\documentclass[apj]{emulateapj}
\usepackage{multirow}
\usepackage{graphics}
\usepackage{rotating}
\usepackage{enumitem}
\usepackage{amsmath}
\usepackage{txfonts}
\usepackage{hyperref}
\usepackage{setspace}
\usepackage{float}
\usepackage{tablefootnote}
\usepackage{footnote}
\usepackage{threeparttable}
\usepackage[mediumspace,mediumqspace,Grey,squaren]{SIunits}

\hypersetup{
			hyperfootnotes=true,
			colorlinks=true,
			linkcolor=blue,
			citecolor=blue,
			urlcolor=blue,
			breaklinks=false,
}

\shorttitle{The central UV analysis of Green Peas and LBAs}
\shortauthors{Kim et al.}

\begin{document}

\title{The Importance of Star Formation Intensity in Ly$\alpha$ Escape From Green Pea Galaxies and Lyman Break Galaxy Analogs}
\author{Keunho Kim\altaffilmark{1}, Sangeeta Malhotra\altaffilmark{2,1}, James E. Rhoads\altaffilmark{2,1}, Nathaniel R. Butler\altaffilmark{1}, and Huan Yang\altaffilmark{3}}

\altaffiltext{1}{School of Earth $\&$ Space Exploration, Arizona State University, Tempe, AZ 85287, USA; Keunho.Kim@asu.edu}
\altaffiltext{2}{NASA Goddard Space Flight Center, Greenbelt, MD 20770, USA}
\altaffiltext{3}{Las Campanas Observatory, Carnegie Institution for Science, Chile}

\def\OI{[\mbox{O\,{\sc i}}]~$\lambda 6300$}
\def\OIII{[\mbox{O\,{\sc iii}}]~$\lambda 5007$}
\def\OIIIs{[\mbox{O\,{\sc iii}}]~$\lambda 4363$}
\def\OIIIab{[\mbox{O\,{\sc iii}}]$\lambda\lambda 4959,5007$}
\def\SIIab{[\mbox{S\,{\sc ii}}]~$\lambda\lambda 6717,6731$}
\def\SII{[\mbox{S\,{\sc ii}}]~$\lambda \lambda 6717,6731$}
\def\NII{[\mbox{N\,{\sc ii}}]~$\lambda 6584$}
\def\NIIb{[\mbox{N\,{\sc ii}}]~$\lambda 6584$}
\def\NIIa{[\mbox{N\,{\sc ii}}]~$\lambda 6548$}
\def\NI{[\mbox{N\,{\sc i}}]~$\lambda \lambda 5198,5200$}

\def\OIIa{[\mbox{O{\sc ii}}]~$\lambda 3726$}
\def\OIIb{[\mbox{O{\sc ii}}]~$\lambda 3729$}
\def\NeIIIa{[\mbox{Ne{\sc iii}}]~$\lambda 3869$}
\def\NeIIIb{[\mbox{Ne{\sc iii}}]~$\lambda 3967$}
\def\OIIIa{[\mbox{O{\sc iii}}]~$\lambda 4959$}
\def\OIIIb{[\mbox{O{\sc iii}}]~$\lambda 5007$}
\def\HeII{{He{\sc ii}}~$\lambda 4686$}
\def\ArIVa{[\mbox{Ar{\sc iv}}]~$\lambda 4711$}
\def\ArIVb{[\mbox{Ar{\sc iv}}]~$\lambda 4740$}
\def\NIa{[\mbox{N{\sc i}}]~$\lambda 5198$}
\def\NIb{[\mbox{N{\sc i}}]~$\lambda 5200$}
\def\HeI{{He{\sc i}}~$\lambda 5876$}
\def\OI{[\mbox{O{\sc i}}]~$\lambda 6300$}
\def\OIb{[\mbox{O{\sc i}}]~$\lambda 6364$}
\def\SIIa{[\mbox{S{\sc ii}}]~$\lambda 6716$}
\def\SIIb{[\mbox{S{\sc ii}}]~$\lambda 6731$}
\def\ArIII{[\mbox{Ar{\sc iii}}]~$\lambda 7136$}

\def\Ha{{H$\alpha\,$}}
\def\Hb{{H$\beta\,$}}

\def\NIIHa{[\mbox{N\,{\sc ii}}]/H$\alpha$}
\def\SIIHa{[\mbox{S\,{\sc ii}}]/H$\alpha$}
\def\OIHa{[\mbox{O\,{\sc i}}]/H$\alpha$}
\def\OIIIHb{[\mbox{O\,{\sc iii}}]/H$\beta$}

\def\Ebmv{E($B-V$)}
\def\LOIII{$L[\mbox{O\,{\sc iii}}]$}
\def\Ledd{${L/L_{\rm Edd}}$}
\def\LOIIIs4{$L[\mbox{O\,{\sc iii}}]$/$\sigma^4$}
\def\LOIIIMbh{$L[\mbox{O\,{\sc iii}}]$/$M_{\rm BH}$}
\def\Mbh{$M_{\rm BH}$}
\def\Msigma{$M_{\rm BH} - \sigma$}
\def\Ms{$M_{\rm *}$}
\def\Msun{$M_{\odot}$}
\def\Msunyr{$M_{\odot}yr^{-1}$}
\def\bt{$B/T\,$}
\def\btr{$B/T_{\rm r}\,$}

\def\ergs{$~\rm ergs^{-1}$}
\def\kms{${\rm km}~{\rm s}^{-1}$}
\newcommand{\cms}{\mbox{${\rm cm\;s^{-1}}$}}
\newcommand{\pccm}{\mbox{${\rm cm^{-3}}$}}
\newcommand{\sauron}{{\texttt {SAURON}}}
\newcommand{\oasis}{{\texttt {OASIS}}}
\newcommand{\HST}{{\it HST\/}}

\newcommand{\Vg}{$V_{\rm gas}$}
\newcommand{\Sg}{$\sigma_{\rm gas}$}
\newcommand{\eg}{e.g.,}
\newcommand{\ie}{i.e.,}

\newcommand{\gandalf}{{\texttt {gandalf}}}
\newcommand{\fracDeV}{{\texttt {FracDeV}}} 
\newcommand{\ppxf}{{\texttt {pPXF}}}

\newcommand{\sersic}{S\'{e}rsic}
\newcommand{\lumpc}{$L_{\rm bol, 250pc}$}
\newcommand{\lumtot}{$L_{\rm bol, total}$}
\newcommand{\sbpc}{$S_{\rm 250pc}$}
\newcommand{\sbeff}{$S_{\rm eff}$}
\newcommand{\reff}{$R_{\rm eff}$}
\newcommand{\rtotal}{$R_{\rm total}$}
\newcommand{\ewlya}{EW(Ly$\alpha$)}
\newcommand{\fesclya}{$f^{Ly\alpha}_{esc}$}
\newcommand{\lya}{Ly$\alpha$}
\begin{abstract}
We have studied ultraviolet images of 40 Green Pea galaxies and 15 local Lyman Break Galaxy Analogs to understand the relation between \lya\ photon escape and central UV photometric properties. We measured star formation intensity (SFI, star formation rate per unit area) from the central 250 pc region (\sbpc) using COS/NUV images from the \textit{Hubble Space Telescope}. The measured \sbpc\ of our sample Green Peas ranges from 2.3--46 $M_{\sun} \ \rm{year}^{-1} \ \rm{kpc^{-2}}$, with a geometric mean of $15 M_{\sun} \ \rm{year}^{-1} \ \rm{kpc^{-2}}$ and a standard deviation of 0.266 dex, forming a relatively narrow distribution. The Lyman Break Galaxy Analogs show a similarly narrow distribution of \sbpc\ (0.271 dex), though with a larger mean of 28 $M_{\sun} \ \rm{year}^{-1} \ \rm{kpc^{-2}}$. We show that while the \lya\ equivalent width (\ewlya) and the \lya\ escape fraction (\fesclya) are not significantly correlated with the central SFI (\sbpc), both are positively correlated with the ratio of surface brightness to galaxy stellar mass ($S_{\rm 250pc}/M_{\rm star}$), with correlation coefficients ($p$-values) of 0.702 ($1\times 10^{-8}$) and 0.529 ($5\times 10^{-4}$) with \ewlya\ and \fesclya , respectively. These correlations suggest a scenario where intense central star formation can drive a galactic wind in galaxies with relatively shallow gravitational potential wells, thus clearing channels for the escape of \lya\ photons. 

\end{abstract}	


\section{Introduction}
\label{sec:introduction}
Green Pea galaxies are a class of local starburst galaxies that were discovered by the citizen science ``Galaxy Zoo'' project based on the Sloan Digital Sky Survey (SDSS) \citep{lint08}. As inferred from their nickname, their optical color is greenish due to their strong \OIII\ emission line at their redshifts (i.e., $0.1 \lesssim z \lesssim 0.35$), and their morphology seen in SDSS images is mostly compact and unresolved \citep[e.g.,][]{card09}. Studies on Green Peas (GPs) have shown that they are low stellar mass (8 $\lesssim {\rm log}(M_{\rm star}/M_{\sun}) \lesssim 10$) and metal-poor galaxies for their stellar mass with typically low intrinsic extinction ($E(B-V) \lesssim 0.2$) and high $[\mbox{O\,{\sc iii}}]$/$[\mbox{O\,{\sc ii}}]$ ratios, experiencing intense star formation activities (i.e., $10^{-7} {\rm{year^{-1}}} \lesssim $ specific star formation rate (sSFR) $\lesssim 10^{-9} {\rm{year^{-1}}}$) \citep[e.g.,][and references therein]{amor10,izot11,jask13,yang16,yang17}. In particular, their UV properties have shown that majority of GPs are \lya -emitters (LAEs) and some of which have been confirmed as Lyman-continuum (LyC) leakers \citep[e.g.,][]{henr15,yang16,yang17,izot16,izot18b,orli18}.

In the field of cosmology, LyC leakers are important possible contributors for reionizing the early Universe ($z>6$). Therefore, in consideration of the associations between LAEs and LyC leakers \citep[e.g.,][]{verh15,deba16,izot16}, searching for LAEs and understanding the \lya\ escape mechanisms is of astrophysical interest \citep[e.g.,][]{ahn03,verh06,gron16}. An ideal approach for studying LAEs would be directly measuring their physical properties from observation \citep[e.g.,][]{dey98,rhoa00,gawi07}. However, since most of LAEs are observed at high redshift \citep[$z \gtrsim 2$, e.g.,][and reference therein]{song14,shib19}, directly observing them has been challenging mainly due to their observed faintness associated with redshift and the intervening intergalactic medium (IGM) absorption along the line of sight. In this regard, an alternative approach for studying high-$z$ LAEs would be studying the physical properties of \textit{local} analogs of high-$z$ LAEs such as GPs \citep[e.g.,][]{izot11,yang16,yang17}.

Morphologically, it has been reported that LAEs are typically ``compact'', often with multiple clumps in the rest-frame UV continuum (i.e., the effective radius \reff\ $\lesssim 1.5$ $\rm{kpc}$) over a wide range of redshift $0 \lesssim z \lesssim 6$ \cite[e.g.,][and references therein]{bond09,malh12,jian13,paul18,shib19,rito19}. While there is an overall consensus regarding the compact and clumpy morphologies of most of LAEs studied, it does not seem entirely clear how these compact morphologies could be related to the observed \lya\ profiles---and more fundamentally, whether the compact/clumpy morphology of \lya -emitting galaxies is one of the important physical conditions that makes a galaxy a \lya -emitting galaxy \cite[e.g.,][]{malh12,izot18}.

In this context, we investigate the central UV photometric properties of LAEs and continuum-selected Lyman Break Galaxy Analogs (LBAs), and the associations with the observed \lya\ line properties based on GPs and local LBAs \citep[i.e.,][]{heck05}. We utilize COS/NUV acquisition images from the \textit{Hubble Space Telescope} (\textit{HST}) and the measured \lya\ properties from the literature (i.e., Alexandroff et al. 2015; Yang et al. 2017, hereafter Y17). The physical proximity of our sample GPs and LBAs (i.e., $0.1 \lesssim z \lesssim 0.35$) and the high angular resolution (i.e., 0.0235 arcsec $\rm{pixel}^{-1}$) of the COS/NUV images are suitable for studying the spatially-resolved central region properties of GPs and LBAs.

Section \ref{sec:samples and data analysis} describes our galaxy sample and the central star formation intensity measurements. In Section \ref{sec:results}, we present our results. In Section \ref{sec:discussion}, we discuss the implications of these results and summarize our primary conclusions. Throughout this paper, we adopt the AB magnitude system and the $\Lambda$CDM cosmology of ($H_{0}$, $\Omega_{m}$, $\Omega_{\Lambda}$) = (70 $\rm{kms^{-1}}$ $\rm{Mpc^{-1}}$, 0.3, 0.7).

\section{SAMPLES AND DATA ANALYSIS}
\label{sec:samples and data analysis}

\subsection{Green Pea and Lyman Break Galaxy Analog sample}
\label{subsec:Sample selection}
Our Green Pea sample is drawn from the 43 galaxies presented in Y17. As described in that paper, all of the galaxies have been observed with \textit{HST}/COS spectroscopy and the associated NUV imaging through the COS acquisition mode ACQ/IMAGE with MIRRORA except for two galaxies (Green Pea ID 0021+0052 and 0938+5428) that have been observed with the MIRRORB configuration. We adopt additional information (e.g., Green Pea ID, equivalent width of the \lya\ line \ewlya , \lya\ escape fraction \fesclya , and $E(B-V)$) for our sample Green Pea galaxies from Y17. Among the 43 galaxies, we exclude the two galaxies observed with the  MIRRORB configuration in this study. We additionally exclude another galaxy (Green Pea ID 0747+2336) because the galaxy has no \lya \ emission line detected in the COS spectroscopy observation (see Y17 for details). Therefore, our final Green Pea sample consists of 40 GPs.

The Lyman Break Galaxy Analog sample is drawn from  21 galaxies analyzed by \citet{alex15}. From the 21 galaxies, we only selected galaxies observed with the MIRRORA configuration, leaving 15 galaxies. LBA ID, galaxy stellar mass, and \ewlya\ are adopted from \citet{alex15}. $E(B-V)$ value for the Milky Way extinction is obtained from the NASA/IPAC Galactic Dust Reddening and Extinction tool. \Ha\ and \Hb\ fluxes for the Balmer decrement method to derive an internal extinction correction in Section \ref{subsec:Central 250pc SFI measurments} are obtained from the MPA-JHU catalog \citep{brin04,trem04}. We also note that 5 of the 15 LBA sample have been classified as GPs in Y17 (i.e., GP (LBA) ID 0055-0021 (J0055), 0926+4428 (J0926), 1025+3622 (J1025), 1428+1653 (J1428), and 1429+0643 (J1429)), and thus were already included in the 40 sample GPs. Therefore, the net increase in sample size is 10 additional objects classified as LBAs but not GPs.  We note that any statistics quoted for the LBA sample include all 15 LBAs (both the 10 ``pure'' LBAs and the 5 overlap objects.)

For this combined sample of GPs and LBAs we use of the COS/NUV images to derive their central UV photometric properties. The exposure time of the images is typically greater than 100 seconds. The pivot wavelength of the observed NUV filter is 2319.7 $\AA$.

\subsection{Deconvolution and Segmentation Maps}
\label{subsec:Segmentation Maps}
We derive segmentation maps of individual galaxies from the NUV images using an approach based on Haar wavelet decomposition. In order to compare the central properties of the entire sample without bias from redshift-dependent resolution effects, we first deconvolved raw NUV images of galaxies with the COS/NUV PSF image of star P330E taken during the \textit{HST} program 11473\footnote{$\rm{http://www.stsci.edu/hst/cos/documents/isrs/ISR2010}$\_10.pdf}.  Specifically, we utilized the Python-based Richardson-Lucy deconvolution package\footnote{$\rm{https://scikit}$-$\rm{image.org/docs/dev/api/skimage.restoration.html} \\ ~~~~~~~~~~~~~~~~~~~~$\#$\rm{skimage.restoration.richardson}$\_$\rm{lucy}$} to perform the deconvolution.

We then proceeded with a Haar wavelet decomposition, which represents the galaxy image as a weighted sum of (mutually orthogonal) 2D boxcar functions. The denoising procedure discards terms in that sum whose coefficients are not significantly different from zero, given the noise in the data. Our Haar denoising procedure\footnote{$\rm{http://butler.lab.asu.edu/wavelet}\_$denoising} is a 2d generalization of the 1D TIPSH algorithm that \citet{kola97} originally developed for modelling transient light curves.

We also estimated the sky background and subtracted from the processed (i.e., PSF-deconvolved and Haar wavelet denoised) COS/NUV images. The sky is measured by taking the average background values from the four $50 \times 50$ pixel corner regions of the $200 \times 200$ pixel (4.7$'' \ \times \ $4.7$''$) galaxy images.

We then apply the \citet{petr76} method to the processed and sky-subtracted images to derive the galaxy segmentation maps. The Petrosian method identifies the central region of a galaxy by defining a local surface brightness threshold $I_{\rm thresh}$ such that  $I_{\rm thresh} = \eta \ \times \ \bar{I}(I > I_{\rm thresh})$. (That is, the threshold surface brightness is a factor of $\eta$ below the average surface brightness enclosed within a contour having surface brightness $I_{\rm thresh}$. While $I_{\rm thresh}$ is implicitly defined, it is nevertheless a uniquely defined quantity for surface brightness profiles where both local surface brightness and total luminosity remain finite, as physics demands of real galaxies.) The $I_{\rm thresh}$ that satisfies the above Petrosian equation is found by sorting image pixels in descending surface brightness order, and thus the associated contour with $I_{\rm thresh}$ is not in general circular. This method has the advantages of being independent of the redshift of a galaxy, and relatively insensitive to dust reddening. We adopt  $\eta = 0.2$, which is widely used for deriving galaxy segmentation maps \citep[e.g.,][]{shim01}. The derived segmentation maps of galaxies are used for measuring the total bolometric luminosity to compare with the central SFI in Section \ref{subsec:Lyman Alpha properties vs. S_250pc and S_250pc divided by stellar mass}.

\begin{figure}
\centering
\includegraphics[width=0.47\textwidth]{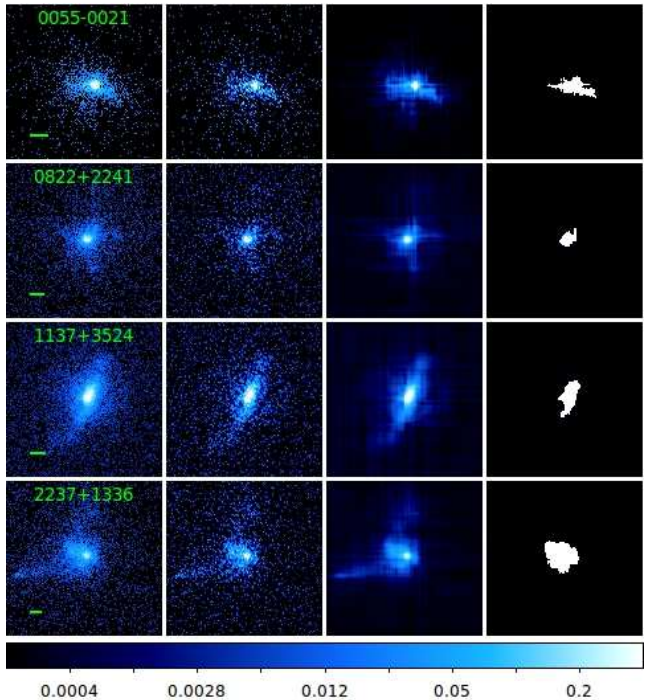}
\caption{Examples of COS/NUV images of some Green Pea samples. Each row shows an individual Green Pea. From left to right, the raw NUV image, the PSF-deconvolved image, the Haar-denoised image, and the segmentation map are displayed, respectively. All images are 3$'' \ \times \ $3$''$ sized. Green Pea ID is marked top middle and a green bar indicates 1 kpc in the raw NUV image. The color bar at the bottom represents flux in the unit of count $\rm{s}^{-1}$.}
\label{fig1}
\end{figure}

\subsection{Star Formation Intensity Measurements from the Central 250 pc Region}
\label{subsec:Central 250pc SFI measurments}

We measure the star formation intensity (SFI, star formation rate per unit area which is equivalent to star formation rate surface density (SFRD)) from the central region of galaxies based on the UV surface brightness in the processed COS/NUV images described in Section \ref{subsec:Segmentation Maps}. Our approach is similar to that employed in \cite{meur97,hath08,malh12}.

We first apply extinction corrections and $k$-correction to estimate the intrinsic UV continuum flux measured at the same rest-frame wavelength for each galaxy. We apply the Milky Way extinction correction following the method of \cite{fitz90} and \cite{fitz99}. We also perform an internal extinction correction adopting the Balmer decrement method and the extinction law from \cite{calz00}. The $k$-correction is performed with respect to the mean rest-frame wavelength of 1877.15 $\angstrom$ in the Green Pea sample. We adopt the intrinsic UV slope of -2 for typical starburst galaxies \cite[e.g.,][]{hath08,malh12}. The galaxy center is set as the brightest pixel in the extinction and $k$-correction processed NUV images.

 We then resample each image to a common pixel scale of 10 pc $\rm{pixel}^{-1}$, using the IRAF ``magnify'' task with a redshift-dependent magnification factor.  At this point, the images of our different sample galaxies have been processed to compensate for nonuniform properties introduced by distance (PSF deconvolution; pixel resampling), redshift (k-correction), and extinction (both foreground and internal).

We then measure the UV luminosity from the central $250\times 250$ pc  region (\lumpc\ in the unit of $L_{\sun}$) together with the associated central SFI (\sbpc\ in the unit of $L_{\sun} \ \rm{kpc^{-2}}$), by directly summing up the flux from $25\times 25$ resampled pixels. We adopt the solar bolometric magnitude of 4.74 \citep{bess98} and the UV to bolometric luminosity ratio ($L_{\rm UV}/L_{\rm bol}$) of 0.33 for typical starbursts \cite[e.g.,][]{meur97,hath08}. We also convert luminosities ($L_{\sun}$) into equivalent star formation rates ($M_{\sun} \ \rm{year}^{-1}$) by using the scale factor $L_{\rm bol}/4.5\times10^{9}L_{\sun} = {\rm{SFR}} / (1\, M_{\sun} \ \rm{year}^{-1})$.  This factor is derived by \cite{meur97} based on starburst population modelling with a solar metallicity and a \cite{salp55} IMF with lower and upper limit masses of $0.1 \ M_\sun$ and $100 \ M_\sun$ respectively.

Examples of raw NUV images, the deconvolved images, the Haar wavelet denoised images, and the derived segmentation maps of some sample galaxies are shown in Figure \ref{fig1}. Also, the measured \sbpc\ and \lumtot\ are provided in Table \ref{tab1}.

\section{Results}
\label{sec:results}

\subsection{Equivalent Width and Escape Fraction of \lya\ Emission versus the Central SFI}
\label{subsec:Lyman Alpha properties vs. S_250pc and S_250pc divided by stellar mass}

We now investigate whether the \lya\ properties of GPs and LBAs are related to their central SFI.\footnote{We note that our analysis with our sample LBAs is limited to their \ewlya, since there is no measured \fesclya\ for our sample LBAs from the literature.} Figure \ref{fig4} shows the relations between \ewlya , \fesclya , and \sbpc . First of all, the measured $\rm{log}$(\sbpc/$M_{\sun} \ \rm{year}^{-1} \ \rm{kpc^{-2}}$) for 40 sample GPs ranges from $\sim 0.37$ to $\sim 1.66$, with a mean $\rm{log}$(\sbpc/$M_{\sun} \ \rm{year}^{-1} \ \rm{kpc^{-2}}$) of 1.17. Compared to approximately two orders of magnitude distributions of \ewlya\ and \fesclya\ of sample GPs, the distribution of their \sbpc\ is narrower by an order of magnitude. It has a standard deviation of 0.266 dex, which corresponds to a factor of $\sim 1.85$. For the 15 sample LBAs (that is, including the 5 LBAs also classified as GPs), their measured $\rm{log}$(\sbpc/$M_{\sun} \ \rm{year}^{-1} \ \rm{kpc^{-2}}$) is typically larger than that of GPs. The mean $\rm{log}$(\sbpc/$M_{\sun} \ \rm{year}^{-1} \ \rm{kpc^{-2}}$) is 1.45 with a standard deviation of 0.271 dex.

We find no significant correlations of either \ewlya\ (panel (a)) or \fesclya\ (panel (b)) with \sbpc, at least for our sample of GPs (i.e., the diamond symbol in the figure) with their relatively narrow dynamic range of \sbpc. The associated Spearman correlation coefficients (hereafter, $r_{\rm{s}}$) ($p$-value) with \ewlya\ and \fesclya\ are only 0.074 (0.7) and -0.027 (0.9), respectively. For comparison, the total bolometric luminosity \lumtot\ shows weak and statistically insignificant anti-correlations with \ewlya\ and \fesclya, with the associated $r_{\rm{s}}$ values ($p$-values) of -0.206 (0.2) and -0.230 (0.2), respectively. Even with the inclusion of LBAs, the correlation between \ewlya\ and \sbpc\ in panel (a) does not seem significant, with the associated $r_{\rm{s}}$ ($p$-value) of -0.079 (0.6).

We also mark the five confirmed LyC leakers from \cite{izot16} among our sample GPs in the figure (i.e., the red-filled diamonds). The  \sbpc \ that we derived for these LyC leakers using NUV-continuum flux is in broad agreement with that derived using H$\beta$ luminosity and the measured NUV-continuum size (i.e., in the unit of scalelength) in \cite{izot16}, matching within a factor of $\sim 2$, except for one galaxy (Green Pea ID 1333+6246) that shows a factor of $\sim 3$ difference between the studies.

\subsection{Examining Specific Star Formation Intensity}
\label{subsec:Intro sSFI}
The specific star formation rate (sSFR), defined as star formation rate normalized by stellar mass, is a powerful summary statistic for the level of star formation in galaxies \citep[e,g.,][]{whit12,kim18}.  Since the power available to drive galactic winds increases with increasing star formation, while the escape velocity for such winds increases with stellar mass, it is reasonable to expect galactic scale outflows to be more common and stronger where sSFR is high.  Very actively star-forming galaxies like  Green Peas and higher redshift \lya\ emitters commonly have sSFR $\ga 10^{-8} \hbox{year}^{-1}$, implying stellar mass doubling times of $< 10^8$ years \citep[e.g.,][]{card09,izot11,fink15,yang17,jian19}.

Next, therefore, we examine the dependences of \ewlya,  \fesclya, and \sbpc\ on both sSFR and also the specific star formation intensity (sSFI), which we define as the SFI divided by total stellar mass. LAEs, including GPs, typically have low stellar mass (8 $\lesssim {\rm log}(M_{\rm star}/M_{\sun}) \lesssim $10) (e.g., Gawiser et al. 2007; Pirzkal et al. 2007; Y17), and show an anti-correlation between stellar mass and \ewlya \ \citep[e.g.,][]{marc19}. This provides a further empirical motivation to investigate the effect of stellar mass on the observed trends between \ewlya , \fesclya , and \sbpc. Figure \ref{fig5} again plots both \ewlya\ and \fesclya, but now as functions of \sbpc\ \textit{divided} by stellar mass (${\rm log} S_{\rm 250pc}/M_{\rm star}$). For our sample GPs, we find that both \ewlya\ and \fesclya\ show significant positive correlations with ${\rm log} S_{\rm 250pc}/M_{\rm star}$. The correlations with ${\rm log} S_{\rm 250pc}/M_{\rm star}$ are stronger than those with sSFR (which are $r_{\rm s} = 0.475$, $p=2\times 10^{-3}$; and $r_{\rm s} = 0.420$, $p= 7\times 10^{-3}$, respectively.) Moreover, when our sample LBAs are also considered in the correlation between \ewlya\ and ${\rm log} S_{\rm 250pc}/M_{\rm star}$ (i.e., panel (a) in the figure), the associated $r_{\rm{s}}$ value shows the most significant correlation coefficient of 0.702 with its $p$-value of $10^{-8}$ among the ones we explored.

All the correlation coefficients are summarized in Table \ref{tab2}.

\begin{figure}
\centering
\includegraphics[width=0.5\textwidth]{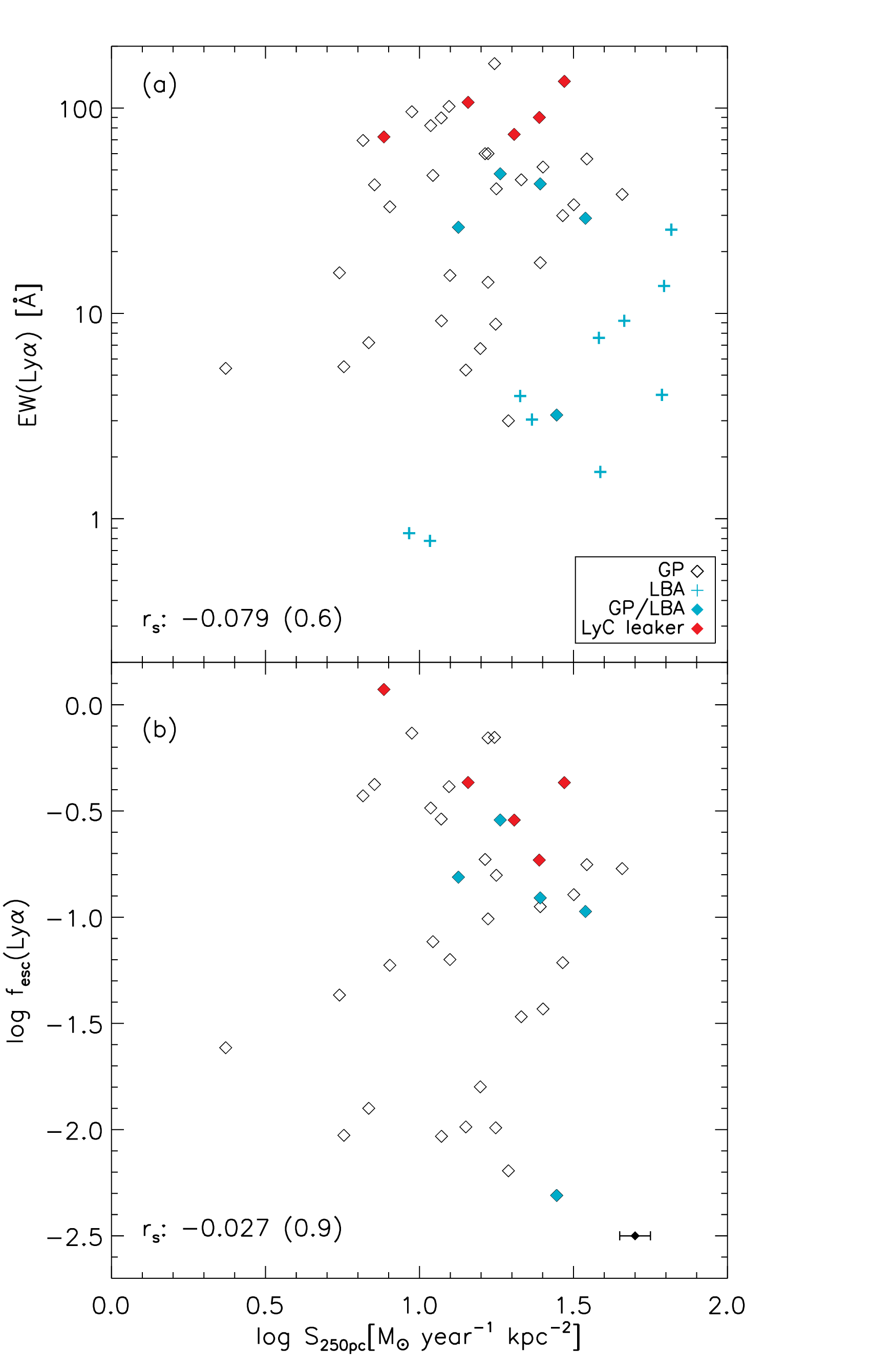}
\caption{\ewlya\ (panel (a)) and \fesclya\ (panel (b)) vs. \sbpc\ for sample GPs and LBAs and sample GPs only, respectively. The associated $r_{\rm{s}}$ values ($p$-values) are shown in bottom left of each panel. In panel (a), the blue crosses are LBAs, while the blue-filled diamonds are the LBAs that are also classified as GPs (See Section \ref{subsec:Sample selection} for details). The red-filled diamonds indicate the five confirmed LyC leakers from \cite{izot16}. The typical measurement error in \sbpc\ due to photon counts and propagation of the errors during the image calibration procedures such as flat-field correction is 0.05 dex, which corresponds to 0.125 error in magnitude. The typical error is marked in bottom right in panel (b). Also, we note that the Milky Way and internal extinction corrections and the $k$-correction performed in Section \ref{subsec:Central 250pc SFI measurments} typically result in 0.29 dex and 0.04 dex corrections in the measured $\rm{log}$(\sbpc/$M_{\sun} \ \rm{year}^{-1} \ \rm{kpc^{-2}}$), respectively. The typical measurement uncertainties of our adopted \ewlya\ and \fesclya\ from Y17 are $\sim 15 \%$ mainly dominated by the systematic error. The similar uncertainties are applied to the measured \ewlya\ from \cite{alex15}. See the text for details.}
\label{fig4}
\end{figure}

\begin{figure}
\centering
\includegraphics[width=0.5\textwidth]{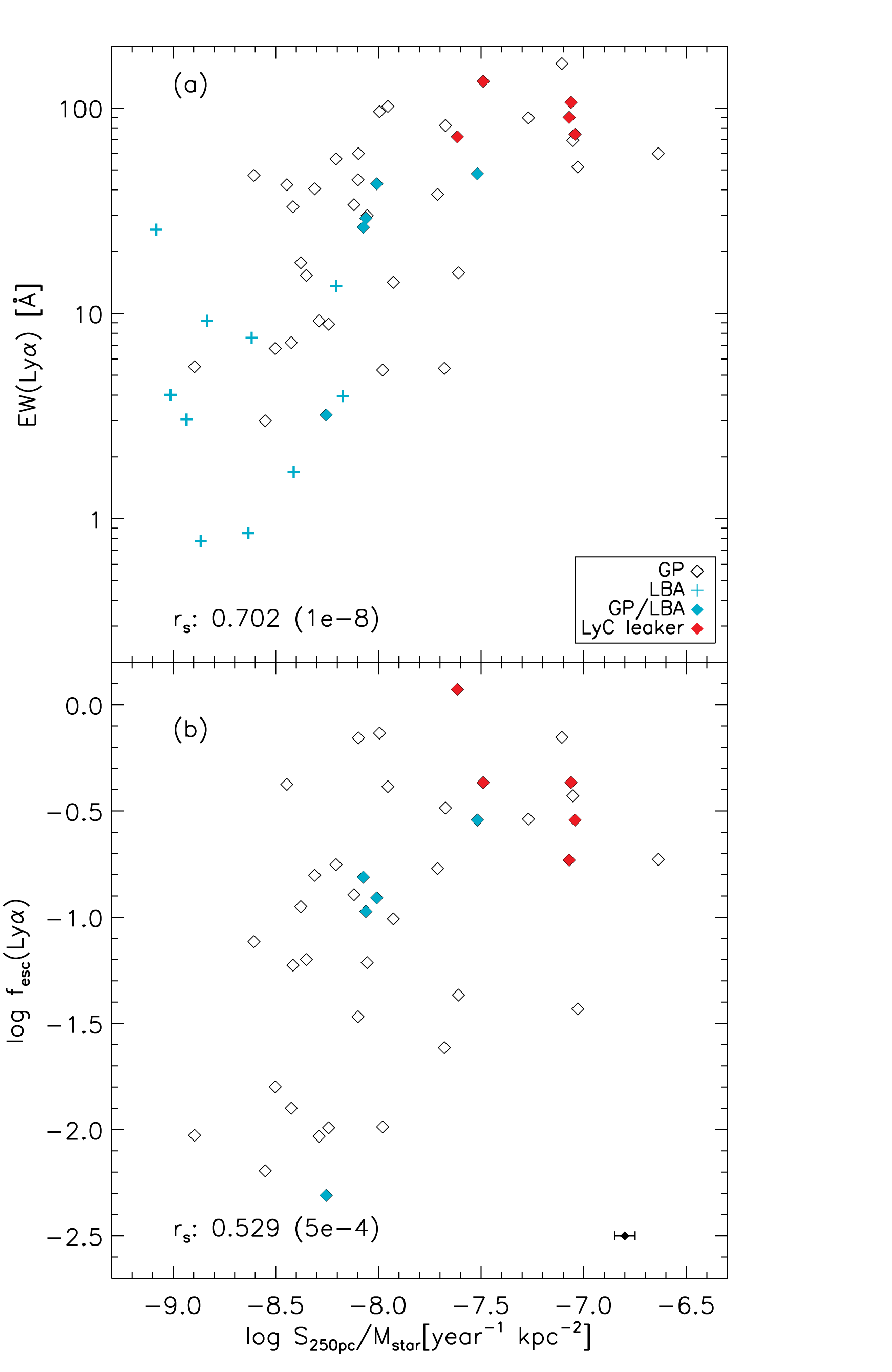}
\caption{Same as Figure \ref{fig4}, but for against ${\rm log} S_{\rm 250pc}/M_{\rm star}$ on x-axis. The more significant positive correlations of \ewlya\ and \fesclya\ with ${\rm log} S_{\rm 250pc}/M_{\rm star}$ than with \sbpc\ are shown.}
\label{fig5}
\end{figure}

As in Figure \ref{fig4}, the five confirmed LyC leakers are marked in the red-filled diamonds in Figure \ref{fig5}. In the parameter space of $S_{\rm 250pc}/M_{\rm star}$, all five LyC leakers have $S_{\rm 250pc}/M_{\rm star} \gtrsim 10^{-7.7} \ {\rm year}^{-1} \ {\rm kpc}^{-2}$, larger than the median of $10^{-8.1}  \ {\rm year}^{-1} \ {\rm kpc}^{-2}$ of the entire sample distribution.

\section{Discussion and Conclusions}
\label{sec:discussion}

In this study, we have identified a sample of Green Pea and Lyman Break Analog galaxies, which are (respectively) some of the best local analogs of high-$z$ \lya- and Lyman break galaxies. Using this sample, we have measured the central UV SFI (\sbpc) and specific star formation intensity \sbpc/$M_{\rm star}$, and investigated the correlations between these two quantities and \lya\ line strength, as characterized by \ewlya\ and \fesclya. We summarize our primary results and conclusions below.

First, the central UV SFI of GPs and LBAs is approximately  three orders of magnitude larger than typical for normal star-forming galaxies \citep[e.g.,][]{kenn98}.  Moreover, this SFI has a relatively narrow distribution, with $\sigma$ $\sim$ 0.27 dex. And, the typical SFI for LBAs is about double that for GPs.  (Specifically, the mean \sbpc\ is $\sim 28$ $M_{\sun} \ \rm{year}^{-1} \ \rm{kpc^{-2}}$ for LBAs, and 15 $M_{\sun} \ \rm{year}^{-1} \ \rm{kpc^{-2}}$ for GPs).

Such high SFI may blow significant holes in the ISM, where the HI column density ($N_{\rm H{\sc I}}$) becomes low enough that \lya\ photons escape more easily.  We note that most \lya -emitting galaxies have compact morphological properties \cite[e.g.,][]{malh12,alex15,verh17,izot18,rito19}, and that their small sizes allow moderate star formation to produce the comparatively high SFI ($\gtrsim 0.1 M_{\sun} \ \rm{year}^{-1} \ \rm{kpc}^{-2}$, \cite{heck01,izot16}) needed for this mechanism.

High SFI is linked to high central pressure $P_{0}$.  This pressure is most likely dominated by contributions from stellar winds and (possibly) supernovae feedback from the central starburst regions, and may drive galactic outflows \cite[e.g.,][]{heck15,heck16}. (Note, however, that intense star formation in nascent massive star clusters can generate high ambient pressure, but under some conditions may also lead to radiative cooling that reduces the resulting feedback  \cite[e.g.,][]{sili18}). Indeed, all of our sample GPs and LBAs have ${\rm{SFI}} > 0.1 M_{\sun} \ \rm{year}^{-1} \ \rm{kpc}^{-2}$, which is a suggested SFI threshold for galactic-scale outflows \cite[i.e.,][]{heck02}. Adopting the relationship between the effective surface brightness \sbeff\ and $P_{0}$ of $S_{\rm eff}/10^{11} L_{\sun} \ {\rm{kpc}^{-2}} = P_{0}/1.63\times10^{-9} {\rm dy} \ \rm{cm^{-2}}$ (which \cite{meur97} derived from a starburst population model), the \sbpc\ distribution of our Green Pea and LBA samples corresponds to the $P_{0}$ range of $10^{6.1} \ \rm{Kcm}^{-3} \lesssim$ $P_{0}/k_{\rm B}$ $\lesssim 10^{7.54} \ \rm{Kcm}^{-3}$, with a median $P_{0}/k_{\rm B}$ of $10^{6.97} \ \rm{Kcm}^{-3}$, where $k_{\rm B}$ is the Boltzmann constant. Although the details and uncertainties should be considered, the $P_{0}/k_{\rm B}$ distribution derived using \sbpc\ is largely consistent with the gas pressures derived using optical emission lines in a sample of GPs and LBAs \citep[i.e.,][]{jian19}.

However, high SFI alone does not guarantee high \lya\ photon escape, since the direct comparisons between \sbpc , \ewlya , and \fesclya\ do not show significant positive correlations. In particular, our sample LBAs show larger \sbpc\ than our sample GPs do, but do not necessarily show larger \ewlya\ (i.e., Figure \ref{fig4}). The absence of significant correlations between \ewlya\ or \fesclya\ and \sbpc\ suggests that additional physical properties beyond high SFI alone must play a large role in \lya\ photon escape. These could include low gas density associated with the ISM geometry, low stellar mass, low gas-phase metallicity, and so on (e.g., Gawiser et al. 2007; Shibuya et al. 2014; Song et al. 2014; Y17).

The strongest correlations in our study are of \ewlya\ and \fesclya\ with $\hbox{sSFI} \equiv S_{\rm 250pc}/M_{\rm star}$ (Figure \ref{fig5} and Table \ref{tab2}), and not with \sbpc\ alone. This suggests that stellar mass (or related galaxy properties)  plays a role in \lya\ escape, such that galaxies with lower stellar masses have higher \lya\ escape for any particular value of \sbpc. The correlations of \ewlya\ with $S_{\rm 250pc}/M_{\rm star}$ ($r_{\rm{s}} = 0.626$, $p=2\times 10^{-5}$) and \fesclya\ with $S_{\rm 250pc}/M_{\rm star}$ ($r_{\rm{s}} = 0.529$, $p=5\times 10^{-4}$) are stronger than the corresponding correlations with $1/M_{\rm star}$ alone ($r_{\rm s} = 0.544$, $p=3\times 10^{-4}$; and $r_{\rm s} = 0.503$, $p= 1\times 10^{-3}$).  They are likewise stronger than the correlations with sSFR, as examined in Section \ref{subsec:Intro sSFI}.

The correlation between \ewlya\ and sSFI becomes even stronger when we add the LBAs to the sample, with $r_{\rm{s}}= 0.702$ and $p=10^{-8}$.

We suggest \lya\ escape in galaxies that have high sSFI results from an interstellar medium with holes in the  neutral hydrogen distribution, and/or outflow of neutral hydrogen (with consequent reduction of resonant scattering optical depth).  Such ISM geometry and kinematics would be produced by the combination of high pressure in regions of high SFI, which provides a driving force for outflows that clear neutral hydrogen; and low mass, which results in shallow gravitational potentials and makes it easier for winds to effectively remove material from the neighborhood of the star formation. There is a metallicity dependence in the conversion from H$\alpha$ luminosity to SFR, but over the full metallicity range of our sample ($7.7 \lesssim 12+{\rm{log(O/H)}} \lesssim 8.5$) this conversion factor changes only 0.1 dex \citep[c.f., Figure 6 of][]{lee09}.  This is insignificant compared to the 2.5 dex spread we observe in sSFI.

The importance of sSFI in UV photon escape is further demonstrated by a progression in measured sSFI between LyC leakers, GPs, and LBAs. LyC leakers typically show the highest sSFI, GPs the next, and LBAs the lowest (i.e., panel (a) in Figure \ref{fig5}).  Both Lyman continuum and \lya\ escape are enabled by low HI column densities, but Lyman continuum escape tends to require lower columns than \lya\ escape, especially given that resonant scattering effects may enable \lya\ to escape for a wider range of geometries \citep[e.g.,][]{neuf91}.

In conclusion, we have examined the relation between \lya\ emission and multiple tracers of star formation activity.   We find that both high star formation intensity (SFI, defined as star formation rate per area) and high specific star formation rate (sSFR, star formation rate per unit stellar mass) are general properties of Green Peas and Lyman Break Galaxy Analogs, distinguishing them from the broader population of star-forming galaxies. But beyond that, we have demonstrated that the specific star formation intensity (sSFI, defined as SFI per unit stellar mass) is the most powerful predictor of \lya\ emission. This is likely because channels of low HI opacity are more easily cleared in galaxies with high sSFI, due to the interplay between star formation intensity that drives galactic winds and gravitational potential that impedes them.

\begin{table*}[ht]
\centering
\begin{threeparttable}
\caption{The measured \sbpc\ and \lumtot, and adopted \ewlya\ and \fesclya of Sample Galaxies}
\begin{tabular}{llllllll}
\hline \hline
Green Pea ID\tnote{a} & SDSS ObjID\tnote{b} & log\sbpc \tnote{c,d} & log(SFI) \tnote{c,e}  & log\lumtot \tnote{c,f} & log(SFR) \tnote{c,g} & \ewlya \tnote{h} & \fesclya \tnote{i} \\
& & ($\rm{L}_{\sun} \ \rm{kpc^{-2}}$) & ($M_{\sun} \ \rm{year}^{-1} \ \rm{kpc^{-2}}$) & ($\rm{L}_{\sun}$) & ($M_{\sun} \ \rm{year}^{-1}$) & ($\angstrom$) & \\
\hline

0055-0021\tnote{j} & 1237663783666581565 & 11.099 $\pm$ 0.100 & 1.445 $\pm$ 0.100 & 10.411 $\pm$ 0.171 & 0.757 $\pm$ 0.171 & 3.20 & 0.005 \\
0303-0759 & 1237652900231053501 & 10.876 $\pm$ 0.025 & 1.223 $\pm$ 0.025 & 10.124 $\pm$ 0.054 & 0.470 $\pm$ 0.054 & 14.17 & 0.098 \\
0339-0725 & 1237649961383493869 & 10.851 $\pm$ 0.050 & 1.198 $\pm$ 0.050 & 10.808 $\pm$ 0.083 & 1.155 $\pm$ 0.083 & 6.74 & 0.016 \\
0749+3337 & 1237674366992646574 & 10.901 $\pm$ 0.054 & 1.247 $\pm$ 0.054 & 10.962 $\pm$ 0.105 & 1.310 $\pm$ 0.105 & 8.86 & 0.010 \\
0751+1638 & 1237673807042708368 & 10.393 $\pm$ 0.075 & 0.740 $\pm$ 0.075 & 10.149 $\pm$ 0.149 & 0.496 $\pm$ 0.149 & 15.77 & 0.043 \\
0805+0925 & 1237667729656905788 & 10.725 $\pm$ 0.077 & 1.071 $\pm$ 0.077 & 10.582 $\pm$ 0.141 & 0.929 $\pm$ 0.141 & 9.20 & 0.009 \\
0815+2156 & 1237664668421849521 & 10.690 $\pm$ 0.017 & 1.036 $\pm$ 0.017 & 9.793 $\pm$ 0.035 & 0.139 $\pm$ 0.035 & 82.15 & 0.327 \\
0822+2241 & 1237664092897083648 & 11.054 $\pm$ 0.038 & 1.401 $\pm$ 0.038 & 10.437 $\pm$ 0.076 & 0.784 $\pm$ 0.076 & 51.62 & 0.037 \\
0911+1831 & 1237667429018697946 & 11.197 $\pm$ 0.017 & 1.543 $\pm$ 0.017 & 10.621 $\pm$ 0.033 & 0.968 $\pm$ 0.033 & 56.53 & 0.177 \\
0917+3152 & 1237661382232768711 & 11.311 $\pm$ 0.032 & 1.658 $\pm$ 0.032 & 10.760 $\pm$ 0.045 & 1.107 $\pm$ 0.045 & 37.97 & 0.169 \\
0925+1403\tnote{k} & 1237671262812897597 & 11.042 $\pm$ 0.014 & 1.389 $\pm$ 0.014 & 10.500 $\pm$ 0.028 & 0.846 $\pm$ 0.028 & 90.01 &  0.186 \\
0926+4428\tnote{j} & 1237657630590107652 & 10.915 $\pm$ 0.080 & 1.262 $\pm$ 0.080 & 10.328 $\pm$ 0.115 & 0.675 $\pm$ 0.115 & 47.85 &  0.287 \\
0927+1740 & 1237667536393142625 & 10.489 $\pm$ 0.071 & 0.835 $\pm$ 0.071 & 10.479 $\pm$ 0.126 & 0.826 $\pm$ 0.126 & 7.20 & 0.013 \\
1009+2916 & 1237665126921011548 & 10.470 $\pm$ 0.047 & 0.817 $\pm$ 0.047 & 9.884 $\pm$ 0.078 & 0.230 $\pm$ 0.078 & 69.54 & 0.373 \\
1018+4106 & 1237661851459584247 & 10.557 $\pm$ 0.045 & 0.904 $\pm$ 0.045 & 10.101 $\pm$ 0.104 & 0.457 $\pm$ 0.104 & 33.05 & 0.059 \\
1025+3622\tnote{j} & 1237664668435677291 & 10.780 $\pm$ 0.048 & 1.126 $\pm$ 0.048 & 10.433 $\pm$ 0.096 & 0.779 $\pm$ 0.096 & 26.27 &   0.154 \\
1032+2717 & 1237667211592794251 & 10.408 $\pm$ 0.064 & 0.755 $\pm$ 0.064 & 9.990 $\pm$ 0.108 & 0.337 $\pm$ 0.108 & 5.50 & 0.009 \\
1054+5238 & 1237658801495474207 & 11.045 $\pm$ 0.034 & 1.392 $\pm$ 0.034 & 10.705 $\pm$ 0.053 & 1.052 $\pm$ 0.053 & 17.65 & 0.112 \\
1122+6154 & 1237655464839479591 & 10.866 $\pm$ 0.033 & 1.213 $\pm$ 0.033 & 9.957 $\pm$ 0.057 & 0.304 $\pm$ 0.057 & 59.95 & 0.187 \\
1133+6514 & 1237651067351073064 & 10.507 $\pm$ 0.030 & 0.854 $\pm$ 0.030 & 10.334 $\pm$ 0.060 & 0.681 $\pm$ 0.060 & 42.30 & 0.422 \\
1137+3524 & 1237665129613885585 & 10.903 $\pm$ 0.029 & 1.250 $\pm$ 0.029 & 10.483 $\pm$ 0.059 & 0.829 $\pm$ 0.059 & 40.45 & 0.157 \\
1152+3400\tnote{k} & 1237665127467647162 & 10.961 $\pm$ 0.021 & 1.308 $\pm$ 0.021 & 10.671 $\pm$ 0.039 & 1.018 $\pm$ 0.039 & 74.45 & 0.287 \\
1205+2620 & 1237667321644908846 & 10.942 $\pm$ 0.039 & 1.289 $\pm$ 0.039 & 10.422 $\pm$ 0.073 & 0.769 $\pm$ 0.073 & 3.00 & 0.006 \\
1219+1526 & 1237661070336852109 & 10.897 $\pm$ 0.023 & 1.244 $\pm$ 0.023 & 10.243 $\pm$ 0.036 & 0.590 $\pm$ 0.036 & 164.55 & 0.702 \\
1244+0216 & 1237671266571387104 & 10.697 $\pm$ 0.021 & 1.044 $\pm$ 0.021 & 10.558 $\pm$ 0.065 & 0.904 $\pm$ 0.065 & 46.98 & 0.077 \\
1249+1234 & 1237661817096962164 & 10.749 $\pm$ 0.019 & 1.096 $\pm$ 0.019 & 10.404 $\pm$ 0.047 & 0.751 $\pm$ 0.047 & 101.82 & 0.412 \\
1333+6246\tnote{k} & 1237651249891967264 & 10.538 $\pm$ 0.017 & 0.885 $\pm$ 0.017 & 10.176 $\pm$ 0.047 & 0.523 $\pm$ 0.047 & 72.34 & 1.180 \\
1339+1516 & 1237664292084318332 & 10.984 $\pm$ 0.037 & 1.331 $\pm$ 0.037 & 10.091 $\pm$ 0.054 & 0.438 $\pm$ 0.054 & 44.74 & 0.034 \\
1424+4217 & 1237661360765730849 & 10.724 $\pm$ 0.033 & 1.071 $\pm$ 0.033 & 10.101 $\pm$ 0.069 & 0.448 $\pm$ 0.069 & 89.53 & 0.290 \\
1428+1653\tnote{j} & 1237668297680683015 & 11.192 $\pm$ 0.032 & 1.539 $\pm$ 0.032 & 10.761 $\pm$ 0.070 & 1.108 $\pm$ 0.070 & 29.07 &  0.106 \\
1429+0643\tnote{j} & 1237662268069511204 & 11.045 $\pm$ 0.073 & 1.392 $\pm$ 0.073 & 10.459 $\pm$ 0.094 & 0.806 $\pm$ 0.094 & 42.75 &  0.123 \\
1440+4619 & 1237662301362978958 & 11.154 $\pm$ 0.040 & 1.501 $\pm$ 0.040 & 10.815 $\pm$ 0.068 & 1.162 $\pm$ 0.068 & 33.82 & 0.128 \\
1442-0209\tnote{k} & 1237655498671849789 & 11.124 $\pm$ 0.018 & 1.470 $\pm$ 0.018 & 10.416 $\pm$ 0.042 & 0.763 $\pm$ 0.042 & 134.90 & 0.430 \\
1454+4528 & 1237662301900964026 & 11.118 $\pm$ 0.036 & 1.465 $\pm$ 0.036 & 10.496 $\pm$ 0.069 & 0.843 $\pm$ 0.069 & 29.95 & 0.061 \\
1457+2232 & 1237665549967294628 & 10.804 $\pm$ 0.019 & 1.151 $\pm$ 0.019 & 9.987 $\pm$ 0.044 & 0.334 $\pm$ 0.044 & 5.30 & 0.010 \\
1503+3644\tnote{k} & 1237661872417407304 & 10.811 $\pm$ 0.014 & 1.158 $\pm$ 0.014 & 10.262 $\pm$ 0.025 & 0.609 $\pm$ 0.025 & 106.61 &  0.431 \\
1514+3852 & 1237661362380734819 & 10.876 $\pm$ 0.031 & 1.223 $\pm$ 0.031 & 10.272 $\pm$ 0.051 & 0.619 $\pm$ 0.051 & 60.00 & 0.698 \\
1543+3446 & 1237662336790036887 & 10.024 $\pm$ 0.080 & 0.371 $\pm$ 0.080 & 9.654 $\pm$ 0.141 & 0.001 $\pm$ 0.141 & 5.40 & 0.024 \\
1559+0841 & 1237662636912280219 & 10.628 $\pm$ 0.039 & 0.975 $\pm$ 0.039 & 9.936 $\pm$ 0.068 & 0.283 $\pm$ 0.068 & 95.96 & 0.735 \\
2237+1336 & 1237656495641788638 & 10.752 $\pm$ 0.054 & 1.099 $\pm$ 0.054 & 10.758 $\pm$ 0.124 & 1.104 $\pm$ 0.124 & 15.31 & 0.063 \\

\hline \hline 
LBA ID\tnote{l} &  &  &  &  &  &  & \\
\hline
J0150 & 1237649918971084879 & 11.019 $\pm$ 0.070 & 1.365 $\pm$ 0.070 & 10.431 $\pm$ 0.145 & 0.778 $\pm$ 0.145 & 3.04 & ... \\
J0213 & 1237649919510446221 & 11.318 $\pm$ 0.073 & 1.665 $\pm$ 0.073 & 10.317 $\pm$ 0.092 & 0.664 $\pm$ 0.092 & 9.20 & ... \\
J0921 & 1237657242433486943 & 11.440 $\pm$ 0.021 & 1.787 $\pm$ 0.021 & 10.885 $\pm$ 0.041 & 1.232 $\pm$ 0.041 & 4.01 & ... \\
J2103 & 1237652598489153748 & 11.471 $\pm$ 0.134 & 1.818 $\pm$ 0.134 & 10.517 $\pm$ 0.147 & 0.864 $\pm$ 0.147 & 25.56 & ...  \\
J1112 & 1237657591929831540 & 11.236 $\pm$ 0.058 & 1.582 $\pm$ 0.058 & 10.683 $\pm$ 0.082 & 1.030 $\pm$ 0.082 & 7.60 & ... \\
J1113 & 1237667212133728444 & 10.620 $\pm$ 0.033 & 0.966 $\pm$ 0.033 & 10.290 $\pm$ 0.133 & 0.637 $\pm$ 0.133 & 0.85 & ... \\
J1144 & 1237662225675124894 & 10.687 $\pm$ 0.049 & 1.034 $\pm$ 0.049 & 10.573 $\pm$ 0.109 & 0.919 $\pm$ 0.109 & 0.78 & ... \\
J1416 & 1237662528992378986 & 11.240 $\pm$ 0.094 & 1.587 $\pm$ 0.094 & 10.504 $\pm$ 0.101 & 0.851 $\pm$ 0.101 & 1.69 & ... \\
J1521 & 1237662264860344485 & 10.980 $\pm$ 0.072 & 1.327 $\pm$ 0.072 & 10.456 $\pm$ 0.095 & 0.802 $\pm$ 0.095 & 3.96 & ... \\
J1612 & 1237662637450592299 & 11.447 $\pm$ 0.071 & 1.794 $\pm$ 0.071 & 10.857 $\pm$ 0.088 & 1.204 $\pm$ 0.088 & 13.60 & ... \\

\hline \hline \\
\label{tab1}

\end{tabular}
{\small
\begin{tablenotes}
\item[a]  The table footnotes for Table 1 are presented below Table 2 due to page space.
\end{tablenotes}
}

\end{threeparttable}
\end{table*}

\begin{table*}[ht]
\centering
\begin{threeparttable}
\caption{The Spearman correlation coefficients between parameters, and the associated probability values}
\begin{tabular}{lllll}
\hline \hline
 & \multicolumn{2}{l}{\ewlya} & \multicolumn{1}{r}{\fesclya} & \\
 Parameter & $r_{\rm s}$ & $p$  & $r_{\rm s}$ & $p$ \\
\hline
GP only &  & & \\
\hline
SFI $\equiv$ \sbpc & 0.074 &0.7  & -0.027 & 0.9 \\
\lumtot & -0.206 & $0.2$  & -0.229 & 0.2  \\
sSFR $\equiv$ SFR $/ M_{\rm star}$ & 0.475 & $2\times 10^{-3}$  & 0.420 & $7\times 10^{-3}$  \\
$1/M_{\rm star}$ & 0.544 & $3\times 10^{-4}$  & 0.503 & $1\times 10^{-3}$ \\
sSFI $\equiv$ \sbpc$/M_{\rm star}$ & 0.626 & $2\times 10^{-5}$ & 0.529 & $5\times 10^{-4}$ \\
\hline \hline
GP+LBA &  &  & \\
\hline
SFI $\equiv$ \sbpc & -0.079 & 0.6  & ...  &  ... \\
\lumtot & -0.272 & $0.06$  & ... & ...  \\
sSFR $\equiv$ SFR $/ M_{\rm star}$ & 0.617 & $2\times 10^{-6}$  & ... & ...  \\
$1/M_{\rm star}$ & 0.645 & $4\times 10^{-7}$  & ... & ... \\
sSFI $\equiv$ \sbpc$/M_{\rm star}$ & 0.702 & $1\times 10^{-8}$ & ... & ... \\

\hline \hline \\

\label{tab2}
\end{tabular}

{\small
\begin{tablenotes}
\item[a]  Green Pea IDs match those in Yang et al 2017.
\item[b] SDSS DR14 BestObjID.
\item[c] The associated errors are flux measurement uncertainties based on photon counting statistics (i.e., Poisson statistics) and propagation of the errors during the image calibration procedures such as flat-field correction. Resulting errors are typically 0.05 dex and 0.081 dex in \sbpc\ and \lumtot, respectively. Additional sources of measurement uncertainty come from the UV-continuum extinction corrections and the $k$-correction described in Section \ref{subsec:Central 250pc SFI measurments}. These corrections are typically 0.29 dex and 0.04 dex, respectively, with uncertainties that are considerably smaller than the corrections but still potentially significant.
\item[d] The measured \sbpc\ in the unit of $\rm{L}_{\sun} \ \rm{kpc^{-2}}$. See Section \ref{subsec:Central 250pc SFI measurments} for details.
\item[e] The measured star formation intensity (SFI) in the unit of $M_{\sun} \ \rm{year}^{-1} \ \rm{kpc^{-2}}$, which is converted from $\rm{L}_{\sun} \ \rm{kpc^{-2}}$ into equivalent star formation rate surface density (SFRD). See Section \ref{subsec:Central 250pc SFI measurments} for details.
\item[f] The measured \lumtot\ in the unit of $\rm{L}_{\sun}$. See Section \ref{subsec:Central 250pc SFI measurments} for details.
\item[g] The measured star formation rate in the unit of $M_{\sun} \ \rm{year}^{-1}$, which is converted from $\rm{L}_{\sun}$ into equivalent star formation rate (SFR). See Section \ref{subsec:Central 250pc SFI measurments} for details.
\item[h] Equivalent Width of \lya\ emission line. \ewlya\ is measured in Yang et al. 2017 and \cite{alex15} for the sample Green Peas and LBAs, respectively. The typical measurement uncertainties are $\sim$ 15 $\%$ mainly dominated by the systematic error.
\item[i] \lya\ Escape fraction measured in Yang et al. 2017. The typical measurement uncertainties are $\sim$ 15 $\%$ mainly dominated by the systematic error.

\item[j] Green Peas that are also classified as Lyman Break Galaxy Analogs by \cite{alex15}.
\item[k] Confirmed Lyman continuum leakers identified by \cite{izot16}.
\item[l] Lyman Break Galaxy Analog IDs match those in \cite{alex15}.
\end{tablenotes}

}
\end{threeparttable}
\end{table*}

\acknowledgments
We thank the anonymous referee for constructive comments that improved the manuscript. K.K. thanks Gerhardt Meurer, Sanchayeeta Borthakur, and Rolf Jansen for helpful discussions. We thank the US National Science Foundation for its financial support through grant AST-1518057. This work has been supported by HST-GO-13779 from STScI, which is operated by the Association of Universities for Research in Astronomy, Inc., for NASA under contract NAS 5-26555.
\clearpage


\clearpage

\begin{thebibliography}{}

\bibitem[\protect\citeauthoryear{Ahn et al.}{2003}]{ahn03} Ahn, S.-H., Lee, H.-W., \& Lee, H.~M.\ 2003, \mnras, 340, 863

\bibitem[\protect\citeauthoryear{Alexandroff et al.}{2015}]{alex15} Alexandroff, R.~M., Heckman, T.~M., Borthakur, S., et al.\ 2015, \apj, 810, 104

\bibitem[\protect\citeauthoryear{Amor{\'\i}n et al.}{2010}]{amor10} Amor{\'\i}n, R.~O., P{\'e}rez-Montero, E., \& V{\'\i}lchez, J.~M.\ 2010, \apj, 715, L128

\bibitem[\protect\citeauthoryear{Bessell et al.}{1998}]{bess98} Bessell, M.~S., Castelli, F., \& Plez, B.\ 1998, \aap, 333, 231 

\bibitem[\protect\citeauthoryear{Bond et al.}{2009}]{bond09} Bond, N.~A., Gawiser, E., Gronwall, C., et al.\ 2009, The Astrophysical Journal, 705, 639

\bibitem[\protect\citeauthoryear{Brinchmann et al.}{2004}]{brin04} Brinchmann, J., Charlot, S., White, S.~D.~M., et al.\ 2004, \mnras, 351, 1151

\bibitem[\protect\citeauthoryear{Calzetti et al.}{2000}]{calz00} Calzetti, D., Armus, L., Bohlin, R.~C., et al.\ 2000, \apj, 533, 682 

\bibitem[\protect\citeauthoryear{Cardamone et al.}{2009}]{card09} Cardamone, C., Schawinski, K., Sarzi, M., et al.\ 2009, \mnras, 399, 1191

\bibitem[\protect\citeauthoryear{de Barros et al.}{2016}]{deba16} de Barros, S., Vanzella, E., Amor{\'\i}n, R., et al.\ 2016, Astronomy and Astrophysics, 585, A51

\bibitem[\protect\citeauthoryear{Dey et al.}{1998}]{dey98} Dey, A., Spinrad, H., Stern, D., et al.\ 1998, The Astrophysical Journal, 498, L93

\bibitem[\protect\citeauthoryear{Finkelstein et al.}{2015}]{fink15} Finkelstein, K.~D., Finkelstein, S.~L., Tilvi, V., et al.\ 2015, \apj, 813, 78

\bibitem[\protect\citeauthoryear{Fitzpatrick \& Massa}{1990}]{fitz90} Fitzpatrick, E.~L., \& Massa, D.\ 1990, \apjs, 72, 163 

\bibitem[\protect\citeauthoryear{Fitzpatrick}{1999}]{fitz99} Fitzpatrick, E.~L.\ 1999, \pasp, 111, 63 

\bibitem[\protect\citeauthoryear{Gawiser et al.}{2007}]{gawi07} Gawiser, E., Francke, H., Lai, K., et al.\ 2007, \apj, 671, 278

\bibitem[\protect\citeauthoryear{Gronke, \& Dijkstra}{2016}]{gron16} Gronke, M., \& Dijkstra, M.\ 2016, \apj, 826, 14

\bibitem[\protect\citeauthoryear{Hathi et al.}{2008}]{hath08} Hathi, N.~P., Malhotra, S., \& Rhoads, J.~E.\ 2008, \apj, 673, 686 

\bibitem[\protect\citeauthoryear{Heckman}{2001}]{heck01} Heckman, T.~M.\ 2001, Gas and Galaxy Evolution, 345

\bibitem[\protect\citeauthoryear{Heckman}{2002}]{heck02} Heckman, T.~M.\ 2002, Extragalactic Gas at Low Redshift, 292

\bibitem[\protect\citeauthoryear{Heckman et al.}{2005}]{heck05} Heckman, T.~M., Hoopes, C.~G., Seibert, M., et al.\ 2005, \apjl, 619, L35

\bibitem[\protect\citeauthoryear{Heckman et al.}{2015}]{heck15} Heckman, T.~M., Alexandroff, R.~M., Borthakur, S., et al.\ 2015, \apj, 809, 147

\bibitem[\protect\citeauthoryear{Heckman, \& Borthakur}{2016}]{heck16} Heckman, T.~M., \& Borthakur, S.\ 2016, \apj, 822, 9

\bibitem[\protect\citeauthoryear{Henry et al.}{2015}]{henr15} Henry, A., Scarlata, C., Martin, C.~L., et al.\ 2015, \apj, 809, 19

\bibitem[\protect\citeauthoryear{Izotov et al.}{2011}]{izot11} Izotov, Y.~I., Guseva, N.~G., \& Thuan, T.~X.\ 2011, \apj, 728, 161

\bibitem[\protect\citeauthoryear{Izotov et al.}{2016}]{izot16} Izotov, Y.~I., Schaerer, D., Thuan, T.~X., et al.\ 2016, \mnras, 461, 3683

\bibitem[\protect\citeauthoryear{Izotov et al.}{2018a}]{izot18} Izotov, Y.~I., Schaerer, D., Worseck, G., et al.\ 2018a, \mnras, 474, 4514

\bibitem[\protect\citeauthoryear{Izotov et al.}{2018b}]{izot18b} Izotov, Y.~I., Worseck, G., Schaerer, D., et al.\ 2018b, \mnras, 478, 4851

\bibitem[\protect\citeauthoryear{Jaskot, \& Oey}{2013}]{jask13} Jaskot, A.~E., \& Oey, M.~S.\ 2013, \apj, 766, 91

\bibitem[\protect\citeauthoryear{Jiang et al.}{2013}]{jian13} Jiang, L., Egami, E., Fan, X., et al.\ 2013, \apj, 773, 153

\bibitem[\protect\citeauthoryear{Jiang et al.}{2019}]{jian19} Jiang, T., Malhotra, S., Yang, H., et al.\ 2019, \apj, 872, 146

\bibitem[\protect\citeauthoryear{Kim et al.}{2018}]{kim18} Kim, K., Malhotra, S., Rhoads, J.~E., et al.\ 2018, \apj, 867, 118

\bibitem[\protect\citeauthoryear{Kennicutt}{1998}]{kenn98} Kennicutt, R.~C.\ 1998, \apj, 498, 541

\bibitem[\protect\citeauthoryear{Kolaczyk}{1997}]{kola97} Kolaczyk, E.~D.\ 1997, \apj, 483, 340 

\bibitem[\protect\citeauthoryear{Lee et al.}{2009}]{lee09} Lee, J.~C., Gil de Paz, A., Tremonti, C., et al.\ 2009, \apj, 706, 599

\bibitem[\protect\citeauthoryear{Lintott et al.}{2008}]{lint08} Lintott, C.~J., Schawinski, K., Slosar, A., et al.\ 2008, \mnras, 389, 1179

\bibitem[\protect\citeauthoryear{Malhotra et al.}{2012}]{malh12} Malhotra, S., Rhoads, J.~E., Finkelstein, S.~L., et al.\ 2012, \apjl, 750, L36 

\bibitem[\protect\citeauthoryear{Marchi et al.}{2019}]{marc19} Marchi, F., Pentericci, L., Guaita, L., et al.\ 2019, \aap, 631, A19

\bibitem[\protect\citeauthoryear{Meurer et al.}{1997}]{meur97} Meurer, G.~R., Heckman, T.~M., Lehnert, M.~D., Leitherer, C., \& Lowenthal, J.\ 1997, \aj, 114, 54 

\bibitem[\protect\citeauthoryear{Neufeld}{1991}]{neuf91} Neufeld, D.~A.\ 1991, \apjl, 370, L85

\bibitem[\protect\citeauthoryear{Orlitov{\'a} et al.}{2018}]{orli18} Orlitov{\'a}, I., Verhamme, A., Henry, A., et al.\ 2018, \aap, 616, A60

\bibitem[\protect\citeauthoryear{Paulino-Afonso et al.}{2018}]{paul18} Paulino-Afonso, A., Sobral, D., Ribeiro, B., et al.\ 2018, \mnras, 476, 5479

\bibitem[\protect\citeauthoryear{Petrosian}{1976}]{petr76} Petrosian, V.\ 1976, \apjl, 209, L1 

\bibitem[\protect\citeauthoryear{Pirzkal et al.}{2007}]{pirz07} Pirzkal, N., Malhotra, S., Rhoads, J.~E., et al.\ 2007, \apj, 667, 49

\bibitem[\protect\citeauthoryear{Rhoads et al.}{2000}]{rhoa00} Rhoads, J.~E., Malhotra, S., Dey, A., et al.\ 2000, \apj, 545, L85

\bibitem[\protect\citeauthoryear{Ritondale et al.}{2019}]{rito19} Ritondale, E., Auger, M.~W., Vegetti, S., et al.\ 2019, \mnras, 482, 4744

\bibitem[\protect\citeauthoryear{Salpeter}{1955}]{salp55} Salpeter, E.~E.\ 1955, \apj, 121, 161

\bibitem[\protect\citeauthoryear{Shibuya et al.}{2019}]{shib19} Shibuya, T., Ouchi, M., Harikane, Y., et al.\ 2019, \apj, 871, 164

\bibitem[\protect\citeauthoryear{Shibuya et al.}{2014}]{shib14} Shibuya, T., Ouchi, M., Nakajima, K., et al.\ 2014, \apj, 788, 74

\bibitem[\protect\citeauthoryear{Shimasaku et al.}{2001}]{shim01} Shimasaku, K., Fukugita, M., Doi, M., et al.\ 2001, \aj, 122, 1238 

\bibitem[\protect\citeauthoryear{Silich, \& Tenorio-Tagle}{2018}]{sili18} Silich, S., \& Tenorio-Tagle, G.\ 2018, \mnras, 478, 5112

\bibitem[\protect\citeauthoryear{Song et al.}{2014}]{song14} Song, M., Finkelstein, S.~L., Gebhardt, K., et al.\ 2014, \apj, 791, 3

\bibitem[\protect\citeauthoryear{Tremonti et al.}{2004}]{trem04} Tremonti, C.~A., Heckman, T.~M., Kauffmann, G., et al.\ 2004, \apj, 613, 898

\bibitem[\protect\citeauthoryear{Verhamme et al.}{2015}]{verh15} Verhamme, A., Orlitov{\'a}, I., Schaerer, D., et al.\ 2015, Astronomy and Astrophysics, 578, A7

\bibitem[\protect\citeauthoryear{Verhamme et al.}{2017}]{verh17} Verhamme, A., Orlitov{\'a}, I., Schaerer, D., et al.\ 2017, \aap, 597, A13

\bibitem[\protect\citeauthoryear{Verhamme et al.}{2006}]{verh06} Verhamme, A., Schaerer, D., \& Maselli, A.\ 2006, \aap, 460, 397

\bibitem[\protect\citeauthoryear{Whitaker et al.}{2012}]{whit12} Whitaker, K.~E., van Dokkum, P.~G., Brammer, G., et al.\ 2012, \apjl, 754, L29

\bibitem[\protect\citeauthoryear{Yang et al.}{2016}]{yang16} Yang, H., Malhotra, S., Gronke, M., et al.\ 2016, \apj, 820, 130.

\bibitem[\protect\citeauthoryear{Yang et al.}{2017}]{yang17} Yang, H., Malhotra, S., Gronke, M., et al.\ 2017, \apj, 844, 171 



\end{thebibliography}
\end{document}